# Physical Time as Human Time

Ruth E. Kastner[*]

University of Maryland, Department of Philosophy, College Park, MD, USA

**\* Correspondence:**

Ruth E. Kastner

rkastner@umd.edu

**Keywords: Time, Flow of Time, Spacetime, Change, Cosmology, Quantum Substratum**

**Abstract: I dissent from the standard assertion of a "Two Times Problem," in which physical time is taken as being at odds with the human sense of a "flow of time." I provide a brief overview of the case to be made for the contrary view: namely, that physical theory is indeed consistent with a genuine temporal dynamism that takes into account the quantum level in connection with spacetime emergence, the latter being supervenient on specific quantum processes.**

My response to "Physical Time Within Human Time" (Gruber, Block and Montemayor, 2022), will be to dissent from what I will call the Received View (RV) of the topic, which sees a "Two Times Problem" (TTP) in which "physical time" is incompatible with "human time." Two key assumptions underlying the RV are presented in Rovelli and Buonamano (2022):

(i) All elementary mechanical laws of nature that we know are invariant under reversal of the direction of time.

(ii) Relativity is incompatible with an objective notion of a global present.

I take issue with both these assumptions. Firstly, (i) is not necessarily true. Perhaps more precisely, the phrase "that we know" really means "that we are willing to consider," where that means excluding the possibility of real non-unitarity at the quantum level. On the other hand: if there *is* real non-unitarity, then physics gains a temporal direction and irreversibility, which falsifies (i). I have already provided specific reasons to think that we don't need to accept (i) in Kastner (2017) and in Kastner (2022, Chapters 5 and 8).[1]

Regarding (ii), relativity can indeed be compatible with a kind of global present once the quantum level is taken into account as a substratum for an emergent spacetime, such that relativity is no longer assumed to be "the whole story." Instead, relativity can be understood as describing the emergent 3+1 construct only. That emergent construct is a proper subset of the entire physical domain, which encompasses also the quantum substratum (QS), the realm of the global present. In other words, the mistake in (ii) is expecting the present to be located *within* the spacetime manifold itself. This expectation is a byproduct of the naive view that spacetime is a kind of "container" or background for all that exists. Let us call this the "spacetime background" assumption, or STB. In short, my claim is that the STB is just as obvious--and yet *wrong*--as the idea that the Sun goes around the Earth; and the paradigm shift we need is of that order.

Thus, neither of the key assumptions leading to an apparent TTP is obligatory, although they are generally presented as such. Indeed, and with all due respect to those who believe ortherwise: rather than "established science," these assumptions are "established dogmas." Once these inappropriate ground rules are rejected based on more comprehensive physics, there need be no discrepancy between "physical time" and "human time."

While I do not have the space to go into detail here, I have argued that STB is precisely what quantum theory demands of us to question—that in fact, 3+1 spacetime does not exhaust the cosmological reality of our world. Results of this research program are presented in Kastner (2022). Thus I maintain that the STB presupposition is what is illusory, while the standard problematic takes it as a veridical component of the world. My research has made the case that physics could indeed be describing *the same world that we experience* as involving change and dynamism from the standpoint of a factually persistent (at least for some length of proper time) entity. The key ingredient in this new picture is integration of the quantum level in an unorthodox (but, I argue, fruitful) formulation. This formulation contradicts the usual view that a physical system (exemplified by a human being) is "a conglomerate of impermanent events." However, rather than stemming from "a desire not to be ephemeral," this view arises simply from taking quantum physics into account in the ontology, such that it is appropriately integrated with the spacetime level.[2]

---

[1] Physical non-unitarity also provides a solid grounding for the decoherence program, which founders on improper mixtures if there are "really" only unitary processes. This is discussed in Kastner (2020).

As noted above, a key aspect of this treatment is acknowledging that quantum systems are not components of the 3+1 spacetime construct, but rather are precursors to it, as sources of potentiality (see, e.g. Kastner, Kauffman and Epperson, 2018). A quantum system such as hydrogen atom is indeed a persistent entity at the QS level, even as it participates in the creation of spacetime events.[3] The system itself is not reducible to those spacetime events, which are *activities* of the system, distinct from the system itself. Nevertheless, fermionic quantum systems have basic temporal attributes arising from their half-integral spin. This periodicity serves to define proper time and reflects persistence of the quantum system, whose ontological nature is not defined or delimited by any particular spacetime event(s).

Quantum systems can, under suitable and well-quantified conditions, give rise to actualized events, which are the elements of an emergent spacetime construct; thus 3+1 spacetime is supervenient, not fundamental.[4] The events in question are essentially the outcomes of "quantum measurements," and their occurrence (via non-unitarity) breaks the elementary time symmetry referenced in (i). However, the term "measurement" has no anthropomorphic content but simply denotes a physically real non-unitary interaction resulting in the transfer of a real photon. This establishes a pair of linked, *actualized events*: emission and absorption of a photon, while the photon itself is their invariant connection (a null interval). Thus, what we call "spacetime" is nothing more than a set of events (actuals) connected by null intervals: in strict ontological terms, there is neither "time" nor "space" (in any invariant sense) in that manifold. Instead, the metrical notions of "time" and "space" are references by which we describe the occurrence of those events relative to an underlying QS periodicity. Since this ontology includes both actuals and potentialities, what also needs to be relinquished is the tacit commitment to actualism, which underlies the mistake of assuming the spacetime manifold to be the entirety of physical reality.

---

[2] The authors discuss an experiment (Gruber et al., 2020b) in which a subject is wearing a virtual reality apparatus that essentially immerses her in a spacetime map taken as a background against which she can move "back and forth" relative to a time index on a map. However, the user simply experiences the implications of the model as programmed into the VR apparatus. The ability to illustrate the model's implications does not demonstrate that the model describes the world. Note also that the experiment does not test (or refute) the experienced continuity of the experimental subject herself. Thus, it remains open to question whether real systems (like people) are in fact reducible to conglomerates of events in a STB.

[3] The Whiteheadian character of this ontology is noted in Kastner (2019), Chapter 5. It is discussed in further detail in Kastner (2022), Chapter 8. It arises straightforwardly from acknowledging that Hilbert space is not commensurable with 3+1 spacetime and therefore that objects described by Hilbert space quantities cannot exist in spacetime. In other words, if the Hilbert space states of quantum theory refer to physically real objects, then physical ontology cannot properly be delimited by 3+1 spacetime.

In conclusion, the impression that we "exist in spacetime" is ironically a non-veridical VR-type experience, a feature of a kind of "user interface" that serves the purposes of biological organisms. "Spacetime" is a phenomenal, emergent level that should not be mistaken as the fundamental physical level at which we exist, which is the quantum substratum. In this picture, there is no "two times problem"; the physics faithfully reflects our experience of time as a dynamical process of emergence that is harmonious with our existence in the present. Finally, it should be emphasized that this picture is not a reversion to idealism; it is thoroughly realist. Both levels of reality--the actual and the possible--are describable by physical theory and are physically accessible.

---

[4] For details, see Kastner (2022, Chapters 5 and 8; relevant excerpted material is available in preprint form: https://arxiv.org/abs/2101.00712, https://arxiv.org/abs/2103.11245) and also Kastner and Cramer, (2018), Kastner (2018). A specific model of spacetime emergence based on entropic gravity is provided in Schlatter and Kastner (2023). That treatment shows how the process of emergence is Lorentz-invariant.

**Acknowledgments.** I thank Gruber *et al* for their invitation to offer this Comment**,** and two anonymous referees for helpful suggestions.

*The author declares that the research was conducted in the absence of any commercial or financial relationships that could be construed as a potential conflict of interest.*